\documentclass[hyperref,UTF8,aps,prl,twocolumn,superscriptaddress,floatfix]{revtex4}
\usepackage{graphicx}% Include figure files
\usepackage{dcolumn}% Align table columns on decimal point
\usepackage{bm}% bold math
\usepackage{times}% font that prx use
\usepackage{color}
\usepackage{appendix}
\usepackage{hyperref}

\begin{document}

\bibliographystyle{apsrev}
\title{Evolution of superconductivity and antiferromagnetic order in Ba(Fe$_{0.92-x}$Co$_{0.08}$V$_x$)$_2$As$_2$}

\author{Jieming Sheng}
\affiliation{Department of Physics and Beijing Key Laboratory of
Opto-electronic Functional Materials $\&$ Micro-nano Devices,
Renmin University of China, Beijing 100872, P. R. China}
\affiliation{Institute of High Energy Physics, Chinese Academy of
Sciences (CAS), Beijing 100049, China}
\affiliation{Spallation
Neutron Source Science Center, Dongguan 523803, China}
\affiliation{Department of Physics, Southern University of Science
and Technology, Shenzhen 518055, China}

\author{Xingguang Li}
\author{Congkuan Tian}
\affiliation{Department of Physics and Beijing Key Laboratory of
Opto-electronic Functional Materials $\&$ Micro-nano Devices,
Renmin University of China, Beijing 100872, P. R. China}
\author{Jianming Song}
\author{Xin Li}
\author{Guangai Sun}
\affiliation{Key Laboratory of Neutron Physics and Institute of
Nuclear Physics and Chemistry, China Academy of Engineering
Physics, Mianyang 621999, China}
\author{Tianlong Xia}
\author{Jinchen Wang}
\author{Juanjuan Liu}
\author{Daye Xu}
\author{Hongxia Zhang}
\affiliation{Department of Physics and Beijing Key Laboratory of
Opto-electronic Functional Materials $\&$ Micro-nano Devices,
Renmin University of China, Beijing 100872, P. R. China}
\author{Xin Tong}
\author{Wei Luo}
\affiliation{Institute of High Energy Physics, Chinese Academy of
Sciences (CAS), Beijing 100049, China} \affiliation{Spallation
Neutron Source Science Center, Dongguan 523803, China}
\author{Liusuo Wu}
\affiliation{Department of Physics, Southern University of Science
and Technology, Shenzhen 518055, China}
\author{Wei Bao}
\affiliation{Department of Physics and Beijing Key Laboratory of
Opto-electronic Functional Materials $\&$ Micro-nano Devices,
Renmin University of China, Beijing 100872, P. R. China}
\affiliation{Department of Physics, City University of Hong Kong,
Kowloon, Hong Kong SAR}
\author{Peng Cheng}
\affiliation{Department of Physics and Beijing Key Laboratory of
Opto-electronic Functional Materials $\&$ Micro-nano Devices,
Renmin University of China, Beijing 100872, P. R. China}

\date{\today}

\begin{abstract}
The vanadium doping effects on superconductivity and magnetism of
iron pnictides are investigated in
Ba(Fe$_{0.92-x}$Co$_{0.08}$V$_x$)$_2$As$_2$ by transport,
susceptibility and neutron scattering measurements. The doping of
magnetic impurity V causes a fast suppression of superconductivity
with T$_c$ reduced at a rate of 7.4~K/1\%V. On the other hand, the
long-range commensurate $C$-type antiferromagnetic order is
recovered upon the V doping. The value of ordered magnetic moments
of Ba(Fe$_{0.92-x}$Co$_{0.08}$V$_x$)$_2$As$_2$ follows a dome-like
evolution versus doping concentration x. A possible Griffiths-type
antiferromagnetic region of multiple coexisting phases in the
phase diagram of Ba(Fe$_{0.92-x}$Co$_{0.08}$V$_x$)$_2$As$_2$ is
identified, in accordance with previous theoretical predictions
based on a cooperative behavior of the magnetic impurities and the
conduction electrons mediating the Ruderman-Kittel-Kasuya-Yosida
interactions between them.
\end{abstract}
\maketitle

\section{Introduction}
Antiferromagnetic (AFM) correlations are closely related to the
unconventional superconductivity (SC) in Fe-based
superconductors\cite{1,2}. The typical parent compound
BaFe$_2$As$_2$ (122) exhibits a collinear $C$-type AFM order and
orthorhombic lattice distortion below T$_N$$\sim$138~K, and
superconductivity gradually emerges with suppressing both the AFM
and structural transitions with charge doping\cite{3,4}. At the
same time, the evolutions of AFM order with chemical doping
present rich features. For example, the AFM order changes from
long-range commensurate to short-range transversely incommensurate
with Co/Ni electron doping and finally disappears with the
avoidance of magnetic quantum critical point (QCP)\cite{5,6}. On
the other hand, hole doping on the Ba site with alkali metals
could induce a tetragonal magnetic phase with spin-reorientation
and a new double-Q AFM order\cite{7,8}. Moreover hole doping on
the Fe site with magnetic impurities Cr/Mn could generate a
competing G-type AFM order and spin fluctuations\cite{9,10}. The
above observations show the diverse responses of Fe-based
magnetism to different impurities and have been considered as
valuable clues to the puzzle of Fe-based superconducting
mechanism, which has not yet been solved. Exploring new impurity
effect therefore is called for.

The Vanadium impurity doping effect on iron pnictides has been
rarely studied until recently. In a previous work, we found that
vanadium serves as magnetic impurity and a very effective hole
donor for the 122 system\cite{11}. The avoided QCP and spin glass
state which were previously reported in the superconducting phase
of Co/Ni-doped BaFe$_{2}$As$_{2}$ can also be realized in
non-superconducting Ba(Fe$_{1-x}$V$_{x}$)$_{2}$As$_{2}$\cite{11}.
A very recent transport and spectroscopy investigation on V-doped
BaFe$_2$As$_2$ found evidence for coexistence of AFM and local
superconducting regions\cite{12}. These experimental findings make
vanadium an interesting impurity probe for further investigations.
So far the studies of vanadium impurities only focused on the
parent compound BaFe$_2$As$_2$. The influences on the
superconductivity in FeAs-122 system is still unknown.

In this paper, we report the physical properties of
Ba(Fe$_{0.92-x}$Co$_{0.08}$V$_x$)$_2$As$_2$, which explore the
effect of magnetic impurity V on the
Ba(Fe$_{0.92}$Co$_{0.08}$)$_2$As$_2$ superconductor with near
optimal T$_c$$\approx$22~K. Besides a fast suppression of
superconductivity, V doping could tune the short-range
incommensurate AFM order back to a long-range commensurate one.
The onset AFM ordering temperature identified from neutron
scattering measurements can be greatly enhanced ($>$70~K) at some
certain doping concentration, indicating possible new magnetic
phases. Finally the phase diagram of
Ba(Fe$_{0.92-x}$Co$_{0.08}$V$_x$)$_2$As$_2$ is given and the
underlying physics is discussed.

\section{Experimental details}

\begin{figure*}[htbp]
\centering
\includegraphics[width=12cm]{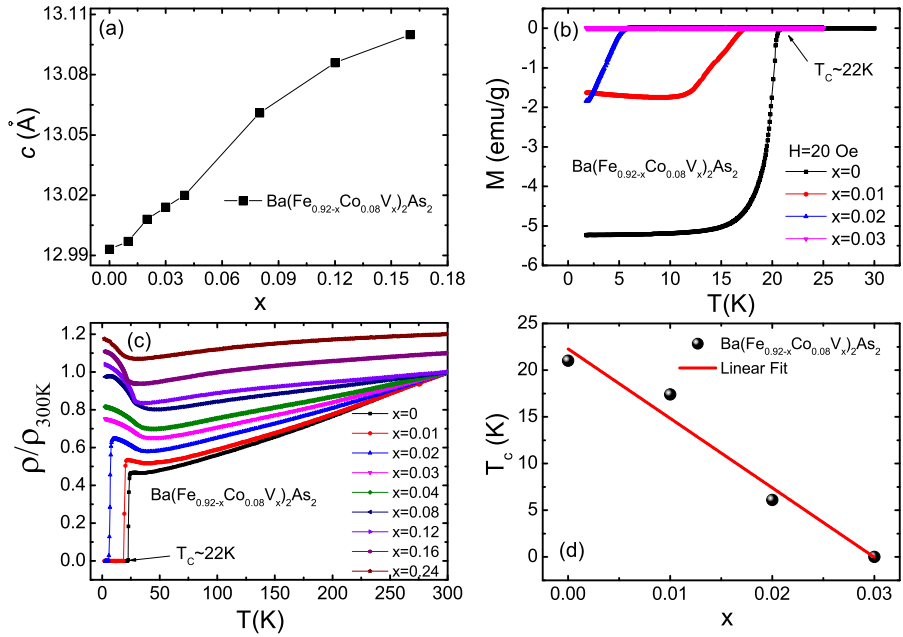}
\caption{(a) Room temperature $c$-axis lattice parameters of the
Ba(Fe$_{0.92-x}$Co$_{0.08}$V$_x$)$_2$As$_2$ series as a function
of nominal V concentration x. (b) Temperature dependence of
zero-field-cooling magnetic susceptibilities for x=0-0.03 samples
under a field of 20~Oe. (c) Temperature dependence of electrical
resistivity for Ba(Fe$_{0.92-x}$Co$_{0.08}$V$_x$)$_2$As$_2$, the
data are normalized to the room temperature value. For x=0.16 and
x=0.24, the data are shifted by a constant value of 0.1 and 0.2
respectively for clarity. (d) The superconducting transition
temperature T$_C$ is plotted as a function of V doping
concentration. The red solid line is the linear fitting result of
the data.}
\end{figure*}

Single crystals of Ba(Fe$_{0.92-x}$Co$_{0.08}$V$_x$)$_2$As$_2$
were grown by self-flux method similar to our previous
report\cite{11}. Crystals with nominal doping between x = 0 and x
= 0.32 are obtained for x-ray, magnetization, electrical transport
and neutron scattering measurements. The x-ray diffraction
patterns were collected from a Bruker D8 Advance x-ray
diffractometer using Cu K$_\alpha$ radiation. The magnetization
measurements of our samples were performed using a Quantum Design
MPMS3. Resistivity measurements were performed on a Quantum Design
physical property measurement system (QD PPMS-14T).

Neutron scattering experiments were carried out on `Xingzhi' cold
neutron triple-axis spectrometer at China advanced research
reactor (CARR)\cite{13} and `Kunpeng' cold neutron triple-axis
spectrometer at China Academy of Engineering Physics (CAEP). The
results below are all reported using the orthorhombic structural
unit cell. For each doping, a single crystal with typical mass of
0.1$\sim$0.2 grams was aligned to the ($H$~0~$L$) scattering
plane. For x=0.12, the crystal was also aligned in the (0~$K$~$L$)
scattering plane, allowing a search for possible incommensurate
AFM phase along the b axis ((0~$K$~0), transverse direction)
similar as that in a previous report\cite{5}.

\section{Results}
Fig. 1(a) presents the doping dependent $c$-axis lattice
parameters of Ba(Fe$_{0.92-x}$Co$_{0.08}$V$_x$)$_2$As$_2$ derived
from the single crystal x-ray data. Vanadium substitution
effectively expands the $c$-axis similar as the effect that
observed in Ba(Fe$_{1-x}$V$_{x}$)$_{2}$As$_{2}$ \cite{11}.

Fig. 1(b) shows the temperature dependence of DC magnetic
susceptibilities for x=0-0.03 samples under a field of 20~Oe. For
x=0, namely Ba(Fe$_{0.92}$Co$_{0.08}$)$_2$As$_2$, the onset
transition of the Meissner effect appears at 22~K. With slight V
doping, both the superconducting transition temperature T$_c$ and
the superconducting shielding volume fraction quickly decrease.
The fast suppression of superconductivity also manifests in the
temperature dependent resistivity data (Fig. 1(c)). T$_c$ is
defined as the onset point of Meissner effect which corresponds
well with the zero-resistivity temperature determined from the R-T
curve. The T$_c$ suppression rate is calculated to be 7.4~K/1\%V
(Fig. 1(d)), similar as that in previous reports about magnetic
impurity doped Fe-based superconductors such as V-doped
LiFeAs\cite{14}, Cr-doped Ba(Fe,Ni)$_2$As$_2$\cite{15} or Mn-doped
Ba$_{0.5}$K$_{0.5}$Fe$_2$As$_2$\cite{16}. On the other hand, the
phase diagram of Ba(Fe,Co)$_{2}$As$_{2}$ has been extensively
studied in early publications\cite{17,5}, consensus has been
reached about the existence of a weak short-range incommensurate
AFM order in the slightly under doped region. According to the
values of lattice parameter, T$_c$ and transport properties, we
can safely put our Ba(Fe$_{0.92}$Co$_{0.08}$)$_2$As$_2$ sample in
this region which have superconductivity at 22~K coexisting with a
short-range incommensurate AFM order below 30~K.

\begin{figure}[htbp]
\centering
\includegraphics[width=0.48\textwidth]{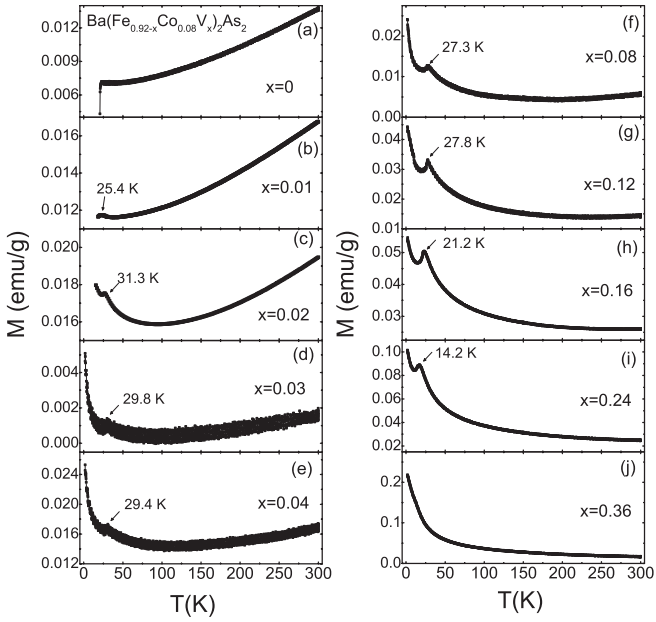}
\caption{(a)-(j) The temperature dependence of normal state DC
susceptibilities for the
Ba(Fe$_{0.92-x}$Co$_{0.08}$V$_x$)$_2$As$_2$ series under magnetic
field of 1~T applied in the H$\parallel$ab direction. The AFM
transition temperatures T$_N$ are determined from the dM/dT
curves.}
\end{figure}

The temperature dependent resistivities of all samples are shown
in Fig. 1(c), an anomaly feature at around 30~K gradually emerges
with increasing x. This anomaly quite resembles the resistivity
anomaly caused by AFM/structural transition in Fe-based 122
materials, which indicates a possible recovery of stronger
magnetic order in the samples. Therefore, as shown in Fig. 2, we
measured the DC magnetic susceptibilities of the normal state for
all samples with H=1~T applied parallel to the $ab$-plane. For
x=0, the short-range incommensurate AFM order is too weak to
generate an AFM feature in the susceptibility data. But for
x$\geq$0.01, the AFM transition features in the M-T curves emerge
and get clearer and sharper with increasing x. The AFM transition
temperature T$_N$ shown in Fig. 2 is defined by the local extremum
of the dM/dT curves, the resistivity anomaly temperatures could
also be determined using the same method and have roughly the same
values as T$_N$ (see the final phase diagram in Fig. 5). It first
increases slightly then decreases monotonously with doping until
could not to be identified at x=0.36.

\begin{figure}[htbp]
\centering
\includegraphics[width=0.48\textwidth]{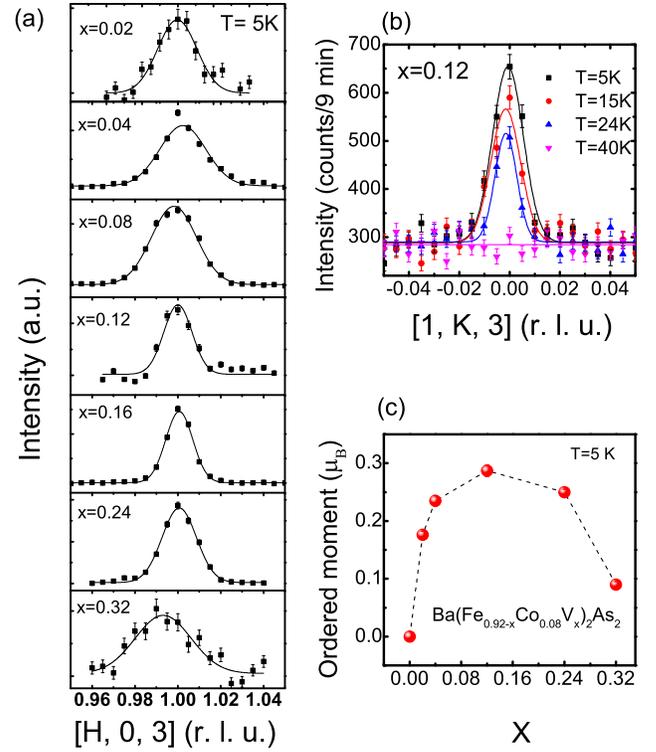}
\caption{(a) Longitudinal ($H$~0~3) scans at Q$_{AFM}$=(1~0~3) and
T=5~K. (b) Transverse (1~$K$~3) scans at Q$_{AFM}$=(1~0~3) at
different temperatures for x=0.12. (c) The ordered magnetic
moments at T=5~K for x=0, 0.02, 0.04, 0.12, 0.24 and 0.32.}
\end{figure}

In order to further clarify the evolution of AFM order in
Ba(Fe$_{0.92-x}$Co$_{0.08}$V$_x$)$_2$As$_2$, we performed elastic
neutron scattering experiments on our samples using `Xingzhi' and
`Kunpeng' cold neutron triple-axis spectrometers. Fig. 3(a)
presents the \textbf{Q} scans for AFM peak Q=(1~0~3) at T=5~K for
x=0.02, x=0.04, x=0.08, x=0.12, x=0.16, x=0.24 and x=0.32 along
\textbf{H} direction. All the raw data of \textbf{Q}-scans are
subtracted by the background above T$_N$, the existence of
magnetic peaks at Q$_{AFM}$=(1~0~3) is evident. The (1~0~1)
magnetic peaks were also collected for all the samples and their
integrated intensities are approximately 45\% of that for (1~0~3)
magnetic peaks. This intensity ratio is consistent with the
calculations from a $C$-type AFM order in BaFe$_2$As$_2$. For some
samples, we also measured at least four nuclear peaks, combined
with the two magnetic peaks, the magnetic ordered moment can be
determined through the refinements using Fullprof software. As
shown in Fig. 3(c), the ordered magnetic moments of
Ba(Fe$_{0.92-x}$Co$_{0.08}$V$_x$)$_2$As$_2$ exhibit a strong
dome-like doping dependent behavior with maximum value of
0.29$\mu_B$, similar evolution of moments was also reported
recently in Cr-doped
BaFe$_{1.9-x}$Ni$_{0.1}$Cr$_x$As$_2$\cite{18}.

To determine the spin-spin correlation length $\xi$, we fit all
the (1~0~3) magnetic peaks by Gaussian function as shown by the
solid lines in Fig. 3(a). The values of $\xi_{ab}$ for V-doped
samples from x=0.02 to x=0.24 are larger than 300$\AA$. For some
samples, such as x=0.04 and x=0.16, the \textbf{L}-scans for
(1~0~3) magnetic peak were also performed and fitted by Gaussian
function (data not shown here), which give a similar result of
$\xi_{c}$$>$300$\AA$. The large values of $\xi$ provide evidence
for a long-range magnetic order has been restored for the V-doped
samples. In addition, the temperature dependence of transverse
\textbf{K}-scan of (1~0~3) magnetic peak for x=0.12 sample were
also measured (shown in Fig. 3(b)). Through the Gaussian fit of
these peaks, the magnetic peak is confirmed to be commensurate for
all measured temperatures.

\begin{figure}[htbp]
\centering
\includegraphics[width=0.48\textwidth]{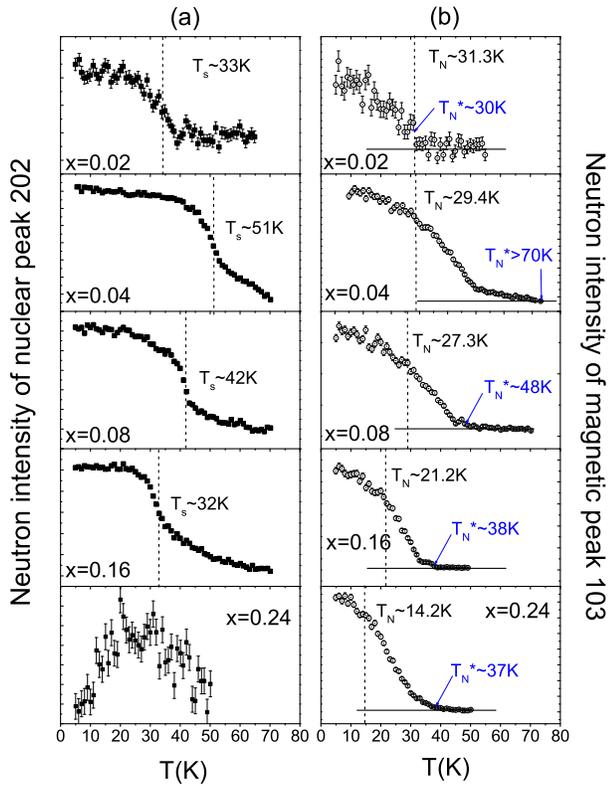}
\caption{Order parameters of structural and magnetic transitions
in Ba(Fe$_{0.92-x}$Co$_{0.08}$V$_x$)$_2$As$_2$: (a) The
temperature dependence of the neutron intensity at Q=(2~0~2). (b)
The temperature dependence of the neutron intensity at
Q$_{AFM}$=(1~0~3). T$_N^*$ is defined as the onset temperature of
(1~0~3) peaks and the dotted lines represent the T$_N$ determined
from the susceptibility data.}
\end{figure}

In order to further investigate the AFM and structural
transitions, the temperature dependence of the intensities of
nuclear Bragg peak (2~0~2) and magnetic Bragg peak (1~0~3) are
studied for Ba(Fe$_{0.92-x}$Co$_{0.08}$V$_x$)$_2$As$_2$,
respectively (shown in Fig. 4). For the tetragonal to orthorhombic
structural transition, the neutron extinction effect from the peak
splitting results in a significant change for the peak intensity
of (2~0~2) around the transition temperature. However, for
x$\geq$0.24, the huge structural factor and strong intensity of
the Bragg peaks make it difficult to figure out the structural
transition via extinction effect\cite{18}. The structural
transition temperatures T$_S$ are roughly determined from the
maximum slope of the change of (2~0~2) intensity as shown in Fig.
4(a). The temperature evolutions of the magnetic order parameters
are illustrated in Fig. 4(b). We notice that the magnetic order
for most components can survive at much higher temperatures
comparing with the T$_N$ determined from susceptibility and
resistivity measurement, especially for x=0.04, which exhibits a
long magnetic intensity tail extending to T=73~K. In Fig.4(b), we
define T$_N^*$ as the onset temperature of the magnetic order from
our neutron diffraction results and the dashed lines mark the
T$_N$ which is determined from the susceptibility data in Fig.2.
Although for x=0.02 there is almost no difference between T$_N^*$
and T$_N$, the deviations of T$_N^*$ and T$_N$ are quite obvious
for x$\geq$0.04. Finally in Fig. 5, the $x-T$ phase diagram of
Ba(Fe$_{0.92-x}$Co$_{0.08}$V$_x$)$_2$As$_2$ is plot based on the
experimental results above. We will discuss the features and
underlying physics concerning to this phase diagram in the next
section.

\section{Discussions and Conclusions}

The above experimental results confirm that the long-range
commensurate magnetic order could be recovered through doping V
into Ba(Fe$_{0.92-x}$Co$_{0.08}$V$_x$)$_2$As$_2$. The ordered
moments of recovered AFM phases display a dome-like evolution
versus x with maximum at x=0.12 (Fig. 3(c)). The similar
recoveries of magnetic phase have been observed in Cr-doped
BaFe$_{1.9-x}$Ni$_{0.1}$As$_2$\cite{18} and K-doped
BaFe$_{2-x}$Co$_x$As$_2$\cite{19}, which seems to be a universal
phenomenon in FeAs-122 system but has never been realized in other
unconventional magnetic superconducting system as far as we know.
In our previous work\cite{11}, V is shown to act as an effective
hole dopant similar as the cases of Cr and K. So a straightforward
explanation about the recovery of magnetic order could be based on
the Fermi surface nesting picture. Namely the introduction of
holes into the electron doped FeAs-122 system compensates the
charges by lowering the chemical potential and reshapes the Fermi
surface. The better condition of Fermi surface nesting stabilizes
the magnetic ordering, and further hole doping breaks the charge
balance and finally diminishes the magnetic order. Although the
Fermi surface nesting picture of the magnetism and
superconductivity in many iron pnictides and iron chalcogenides
materials has been challenged, it provides a good explanation for
our experimental observations. The nature of magnetism in Fe-based
superconductors remains elusive, the above observations of the
charge tunable magnetic orders in FeAs-122 system should provide
insights on the final physical picture.

\begin{figure}[htbp]
\centering
\includegraphics[width=0.48\textwidth]{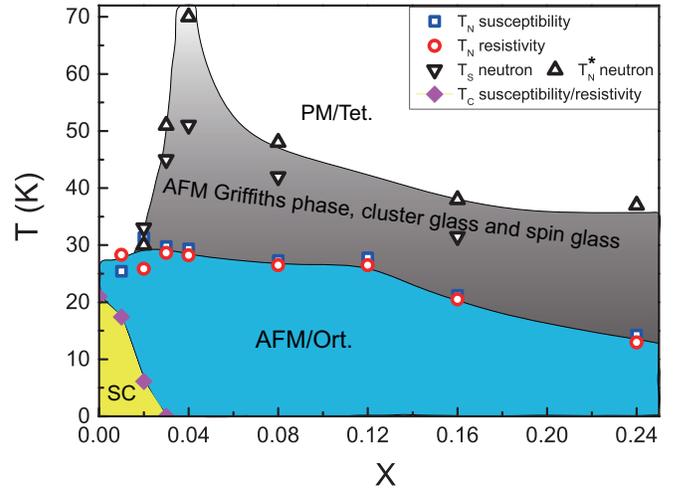}
\caption{$T$-$x$ phase diagram of
Ba(Fe$_{0.92-x}$Co$_{0.08}$V$_x$)$_2$As$_2$ single crystals.}
\end{figure}

As shown in Fig.5, the AFM ordering temperatures T$_N$ determined
from susceptibility and resistivity data are consistent. They are
around 14~K-30~K for all samples and exhibit a weak doping
dependent behavior. However, neutron scattering experiments reveal
that the magnetic peak (1~0~3) has notable intensities at
temperatures much higher than T$_N$ for most samples. Especially
for x=0.04, the intensity of magnetic peak exists even above 70~K,
which is significantly larger than 29.4~K determined from the
susceptibility data. The similar phenomenon has been previously
reported by D. S. Inosov $et$ $al.$ in
Ba(Fe$_{0.88}$Mn$_{0.12}$)$_2$As$_2$ which is considered as an
coexistence of AFM Griffiths phase\cite{20}, AFM cluster glass and
spin glass states in the region between T$_N$ and
T$_N^*$\cite{21}. Normally these states would not generate sharp
transition features in susceptibility and resistivity curves. But
they can make notable slow dynamic contributions to the magnetic
peak intensities which could be detected by neutron scattering
measurements. Later, M. N. Gastiasoro $et$ $al.$ used a realistic
five-band model with standard on-site Coulomb repulsion to study
the magnetic order nucleated by magnetic impurities in iron
pnictides\cite{22}. They found that the magnetic tails of $C$-type
AFM modulations close to the cores of magnetic impurities, may
overlap with neighboring impurities and induce $C$-type AFM order
even above T$_N$ in a clear iron-pnictides system. This provides a
microscopic explanation about the enhancement of $C$-type AFM
ordering temperature in magnetic impurity doped iron pnictides.
Similarly, impurity induced glassy clusters have been observed in
the Fe(Te,Se) systems\cite{23,24,25}. So based on the analysis
above, our data strongly suggest the existence of the AFM
Griffiths regime of multiple coexisting phases between T$_N$ and
T$_N^*$ in the phase diagram of
Ba(Fe$_{0.92-x}$Co$_{0.08}$V$_x$)$_2$As$_2$ as shown in Fig.5.
Besides the similar T$_N$ deviation between susceptibility and
neutron data as reported in
Ba(Fe$_{0.88}$Mn$_{0.12}$)$_2$As$_2$\cite{21}, another evidence is
stated below. M. N. Gastiasoro $et$ $al.$'s calculations\cite{22}
predict that the enhancement of $C$-type AFM order is strongly
doping dependent, namely it occurs and reaches maximum above a
certain doping concentration then gets weakened but always exists
at higher doping (as shown in the Fig.1(e) of reference\cite{22}).
According to the phase diagram in Fig.5, the doping evolution of
enhanced $C$-type AFM ordering temperature T$_N^*$ agrees very
well with the above theoretical prediction. Namely for
Ba(Fe$_{0.92-x}$Co$_{0.08}$V$_x$)$_2$As$_2$, the enhancement of
T$^*_N$ starts at x=0.03 and suddenly reaches maximum at x=0.04,
then gets weakened but always exists from x=0.08 to x=0.24. This
consistency validates the physical picture provided by M. N.
Gastiasoro $et$ $al.$ in explaining the phase diagram of
Ba(Fe$_{0.92-x}$Co$_{0.08}$V$_x$)$_2$As$_2$, which reflects a
cooperative behavior of the magnetic impurities and the conduction
electrons mediating the Ruderman-Kittel-Kasuya-Yosida(RKKY)
interactions between them. So far as we know, our observations
seem to be the first experimental evidence for the theoretical
predictions made by M. N. Gastiasoro $et$ $al.$ about the doping
dependent behavior of magnetic impurities-enhanced AFM order. Most
of the experimental reports about Griffiths-like phases are
concerned with ferromagnetic materials\cite{20}, while the
observations of such phases in antiferromagnets are very rare. So
Ba(Fe$_{0.92-x}$Co$_{0.08}$V$_x$)$_2$As$_2$ could serve as a new
system for further researches on exotic magnetic interactions.

In summary, the phase diagram of
Ba(Fe$_{0.92-x}$Co$_{0.08}$V$_x$)$_2$As$_2$ has been investigated
using x-ray, transport, magnetic susceptibility and neutron
scattering measurements. The vanadium magnetic impurity could
quickly suppress the superconductivity of
Ba(Fe$_{0.92}$Co$_{0.08}$)$_2$As$_2$ and restore a long-range
commensurate $C$-type AFM order. The evolution of AFM ordered
moments exhibits a dome-like behavior with V doping, indicating a
Fermi-surface nesting picture of magnetism in FeAs-122 system. On
the other hand the evolution of doping dependent AFM ordering
temperatures reveals the possible existence of AFM Griffiths
phases, AFM cluster glass and spin glass in
Ba(Fe$_{0.92-x}$Co$_{0.08}$V$_x$)$_2$As$_2$, which also provides
as an experimental evidence for the previous theoretical
prediction based on the RKKY interactions. The above results
demonstrate the rich physics when vanadium magnetic impurities are
introduced into iron pnictides superconductor, which may shed new
light on understanding the Fe-based supercondcuting mechanism.

\section{Acknowledgments}
This work is supported by the National Natural Science Foundation
of China (No. 11227906 and No. 11204373). This work is also
supported by the National Nature Science Foundation of China (No.
11875265), the Scientific Instrument Developing Project of the
Chinese Academy of Sciences (3He based neutron polarization
devices)£¬the Institute of High Energy Physics, and the Chinese
Academy of Science.

\bibliography{Bibtex}

\begin{thebibliography}{25}
\expandafter\ifx\csname natexlab\endcsname\relax\def\natexlab#1{#1}\fi
\expandafter\ifx\csname bibnamefont\endcsname\relax
  \def\bibnamefont#1{#1}\fi
\expandafter\ifx\csname bibfnamefont\endcsname\relax
  \def\bibfnamefont#1{#1}\fi
\expandafter\ifx\csname citenamefont\endcsname\relax
  \def\citenamefont#1{#1}\fi
\expandafter\ifx\csname url\endcsname\relax
  \def\url#1{\texttt{#1}}\fi
\expandafter\ifx\csname urlprefix\endcsname\relax\def\urlprefix{URL }\fi
\providecommand{\bibinfo}[2]{#2}
\providecommand{\eprint}[2][]{\url{#2}}

\bibitem[{\citenamefont{Dai}(2015)}]{1}
\bibinfo{author}{\bibfnamefont{P.}~\bibnamefont{Dai}}, \bibinfo{journal}{Rev.
  Mod. Phys.} \textbf{\bibinfo{volume}{87}}, \bibinfo{pages}{855}
  (\bibinfo{year}{2015}),
  \urlprefix\url{https://dx.doi.org/10.1103/RevModPhys.87.855}.

\bibitem[{\citenamefont{Bao}(2013)}]{2}
\bibinfo{author}{\bibfnamefont{W.}~\bibnamefont{Bao}}, \bibinfo{journal}{Chin.
  Phys. B} \textbf{\bibinfo{volume}{22}}, \bibinfo{pages}{087405}
  (\bibinfo{year}{2013}),
  \urlprefix\url{https://dx.doi.org/10.1088/1674-1056/22/8/087405}.

\bibitem[{\citenamefont{Rotter et~al.}(2008)\citenamefont{Rotter, Tegel, and
  Johrendt}}]{3}
\bibinfo{author}{\bibfnamefont{M.}~\bibnamefont{Rotter}},
  \bibinfo{author}{\bibfnamefont{M.}~\bibnamefont{Tegel}}, \bibnamefont{and}
  \bibinfo{author}{\bibfnamefont{D.}~\bibnamefont{Johrendt}},
  \bibinfo{journal}{Phys. Rev. Lett.} \textbf{\bibinfo{volume}{101}},
  \bibinfo{pages}{107006} (\bibinfo{year}{2008}),
  \urlprefix\url{https://link.aps.org/doi/10.1103/PhysRevLett.101.107006}.

\bibitem[{\citenamefont{Huang et~al.}(2008)\citenamefont{Huang, Qiu, Bao,
  Green, Lynn, Gasparovic, Wu, Wu, and Chen}}]{4}
\bibinfo{author}{\bibfnamefont{Q.}~\bibnamefont{Huang}},
  \bibinfo{author}{\bibfnamefont{Y.}~\bibnamefont{Qiu}},
  \bibinfo{author}{\bibfnamefont{W.}~\bibnamefont{Bao}},
  \bibinfo{author}{\bibfnamefont{M.~A.} \bibnamefont{Green}},
  \bibinfo{author}{\bibfnamefont{J.~W.} \bibnamefont{Lynn}},
  \bibinfo{author}{\bibfnamefont{Y.~C.} \bibnamefont{Gasparovic}},
  \bibinfo{author}{\bibfnamefont{T.}~\bibnamefont{Wu}},
  \bibinfo{author}{\bibfnamefont{G.}~\bibnamefont{Wu}}, \bibnamefont{and}
  \bibinfo{author}{\bibfnamefont{X.~H.} \bibnamefont{Chen}},
  \bibinfo{journal}{Phys. Rev. Lett.} \textbf{\bibinfo{volume}{101}},
  \bibinfo{pages}{257003} (\bibinfo{year}{2008}),
  \urlprefix\url{https://link.aps.org/doi/10.1103/PhysRevLett.101.257003}.

\bibitem[{\citenamefont{Pratt et~al.}(2011)\citenamefont{Pratt, Kim, Kreyssig,
  Lee, Tucker, Thaler, Tian, Zarestky, Bud'ko, Canfield et~al.}}]{5}
\bibinfo{author}{\bibfnamefont{D.~K.} \bibnamefont{Pratt}},
  \bibinfo{author}{\bibfnamefont{M.~G.} \bibnamefont{Kim}},
  \bibinfo{author}{\bibfnamefont{A.}~\bibnamefont{Kreyssig}},
  \bibinfo{author}{\bibfnamefont{Y.~B.} \bibnamefont{Lee}},
  \bibinfo{author}{\bibfnamefont{G.~S.} \bibnamefont{Tucker}},
  \bibinfo{author}{\bibfnamefont{A.}~\bibnamefont{Thaler}},
  \bibinfo{author}{\bibfnamefont{W.}~\bibnamefont{Tian}},
  \bibinfo{author}{\bibfnamefont{J.~L.} \bibnamefont{Zarestky}},
  \bibinfo{author}{\bibfnamefont{S.~L.} \bibnamefont{Bud'ko}},
  \bibinfo{author}{\bibfnamefont{P.~C.} \bibnamefont{Canfield}},
  \bibnamefont{et~al.}, \bibinfo{journal}{Phys. Rev. Lett.}
  \textbf{\bibinfo{volume}{106}}, \bibinfo{pages}{257001}
  (\bibinfo{year}{2011}),
  \urlprefix\url{https://link.aps.org/doi/10.1103/PhysRevLett.106.257001}.

\bibitem[{\citenamefont{Lu et~al.}(2013)\citenamefont{Lu, Gretarsson, Zhang,
  Liu, Luo, Tian, Laver, Yamani, Kim, Nevidomskyy et~al.}}]{6}
\bibinfo{author}{\bibfnamefont{X.}~\bibnamefont{Lu}},
  \bibinfo{author}{\bibfnamefont{H.}~\bibnamefont{Gretarsson}},
  \bibinfo{author}{\bibfnamefont{R.}~\bibnamefont{Zhang}},
  \bibinfo{author}{\bibfnamefont{X.}~\bibnamefont{Liu}},
  \bibinfo{author}{\bibfnamefont{H.}~\bibnamefont{Luo}},
  \bibinfo{author}{\bibfnamefont{W.}~\bibnamefont{Tian}},
  \bibinfo{author}{\bibfnamefont{M.}~\bibnamefont{Laver}},
  \bibinfo{author}{\bibfnamefont{Z.}~\bibnamefont{Yamani}},
  \bibinfo{author}{\bibfnamefont{Y.-J.} \bibnamefont{Kim}},
  \bibinfo{author}{\bibfnamefont{A.~H.} \bibnamefont{Nevidomskyy}},
  \bibnamefont{et~al.}, \bibinfo{journal}{Phys. Rev. Lett.}
  \textbf{\bibinfo{volume}{110}}, \bibinfo{pages}{257001}
  (\bibinfo{year}{2013}),
  \urlprefix\url{https://link.aps.org/doi/10.1103/PhysRevLett.110.257001}.

\bibitem[{\citenamefont{Wa\ss{}er et~al.}(2015)\citenamefont{Wa\ss{}er,
  Schneidewind, Sidis, Wurmehl, Aswartham, B\"uchner, and Braden}}]{7}
\bibinfo{author}{\bibfnamefont{F.}~\bibnamefont{Wa\ss{}er}},
  \bibinfo{author}{\bibfnamefont{A.}~\bibnamefont{Schneidewind}},
  \bibinfo{author}{\bibfnamefont{Y.}~\bibnamefont{Sidis}},
  \bibinfo{author}{\bibfnamefont{S.}~\bibnamefont{Wurmehl}},
  \bibinfo{author}{\bibfnamefont{S.}~\bibnamefont{Aswartham}},
  \bibinfo{author}{\bibfnamefont{B.}~\bibnamefont{B\"uchner}},
  \bibnamefont{and} \bibinfo{author}{\bibfnamefont{M.}~\bibnamefont{Braden}},
  \bibinfo{journal}{Phys. Rev. B} \textbf{\bibinfo{volume}{91}},
  \bibinfo{pages}{060505} (\bibinfo{year}{2015}),
  \urlprefix\url{https://link.aps.org/doi/10.1103/PhysRevB.91.060505}.

\bibitem[{\citenamefont{Allred et~al.}(2016)\citenamefont{Allred, Taddei,
  Bugaris, Krogstad, Lapidus, Chung, Claus, Kanatzidis, Brown, Kang
  et~al.}}]{8}
\bibinfo{author}{\bibfnamefont{J.~M.} \bibnamefont{Allred}},
  \bibinfo{author}{\bibfnamefont{K.~M.} \bibnamefont{Taddei}},
  \bibinfo{author}{\bibfnamefont{D.~E.} \bibnamefont{Bugaris}},
  \bibinfo{author}{\bibfnamefont{M.~J.} \bibnamefont{Krogstad}},
  \bibinfo{author}{\bibfnamefont{S.~H.} \bibnamefont{Lapidus}},
  \bibinfo{author}{\bibfnamefont{D.~Y.} \bibnamefont{Chung}},
  \bibinfo{author}{\bibfnamefont{H.}~\bibnamefont{Claus}},
  \bibinfo{author}{\bibfnamefont{M.~G.} \bibnamefont{Kanatzidis}},
  \bibinfo{author}{\bibfnamefont{D.~E.} \bibnamefont{Brown}},
  \bibinfo{author}{\bibfnamefont{J.}~\bibnamefont{Kang}}, \bibnamefont{et~al.},
  \bibinfo{journal}{Nat. Phys.} \textbf{\bibinfo{volume}{12}},
  \bibinfo{pages}{493} (\bibinfo{year}{2016}),
  \urlprefix\url{https://dx.doi.org/10.1038/NPHYS3629}.

\bibitem[{\citenamefont{Marty et~al.}(2011)\citenamefont{Marty, Christianson,
  Wang, Matsuda, Cao, VanBebber, Zarestky, Singh, Sefat, and Lumsden}}]{9}
\bibinfo{author}{\bibfnamefont{K.}~\bibnamefont{Marty}},
  \bibinfo{author}{\bibfnamefont{A.~D.} \bibnamefont{Christianson}},
  \bibinfo{author}{\bibfnamefont{C.~H.} \bibnamefont{Wang}},
  \bibinfo{author}{\bibfnamefont{M.}~\bibnamefont{Matsuda}},
  \bibinfo{author}{\bibfnamefont{H.}~\bibnamefont{Cao}},
  \bibinfo{author}{\bibfnamefont{L.~H.} \bibnamefont{VanBebber}},
  \bibinfo{author}{\bibfnamefont{J.~L.} \bibnamefont{Zarestky}},
  \bibinfo{author}{\bibfnamefont{D.~J.} \bibnamefont{Singh}},
  \bibinfo{author}{\bibfnamefont{A.~S.} \bibnamefont{Sefat}}, \bibnamefont{and}
  \bibinfo{author}{\bibfnamefont{M.~D.} \bibnamefont{Lumsden}},
  \bibinfo{journal}{Phys. Rev. B} \textbf{\bibinfo{volume}{83}},
  \bibinfo{pages}{060509} (\bibinfo{year}{2011}),
  \urlprefix\url{https://link.aps.org/doi/10.1103/PhysRevB.83.060509}.

\bibitem[{\citenamefont{Tucker et~al.}(2012)\citenamefont{Tucker, Pratt, Kim,
  Ran, Thaler, Granroth, Marty, Tian, Zarestky, Lumsden et~al.}}]{10}
\bibinfo{author}{\bibfnamefont{G.~S.} \bibnamefont{Tucker}},
  \bibinfo{author}{\bibfnamefont{D.~K.} \bibnamefont{Pratt}},
  \bibinfo{author}{\bibfnamefont{M.~G.} \bibnamefont{Kim}},
  \bibinfo{author}{\bibfnamefont{S.}~\bibnamefont{Ran}},
  \bibinfo{author}{\bibfnamefont{A.}~\bibnamefont{Thaler}},
  \bibinfo{author}{\bibfnamefont{G.~E.} \bibnamefont{Granroth}},
  \bibinfo{author}{\bibfnamefont{K.}~\bibnamefont{Marty}},
  \bibinfo{author}{\bibfnamefont{W.}~\bibnamefont{Tian}},
  \bibinfo{author}{\bibfnamefont{J.~L.} \bibnamefont{Zarestky}},
  \bibinfo{author}{\bibfnamefont{M.~D.} \bibnamefont{Lumsden}},
  \bibnamefont{et~al.}, \bibinfo{journal}{Phys. Rev. B}
  \textbf{\bibinfo{volume}{86}}, \bibinfo{pages}{020503}
  (\bibinfo{year}{2012}),
  \urlprefix\url{https://link.aps.org/doi/10.1103/PhysRevB.86.020503}.

\bibitem[{\citenamefont{Li et~al.}(2018)\citenamefont{Li, Sheng, Tian, Wang,
  Xia, Wang, Ye, Tian, Wang, Liu et~al.}}]{11}
\bibinfo{author}{\bibfnamefont{X.-G.} \bibnamefont{Li}},
  \bibinfo{author}{\bibfnamefont{J.-M.} \bibnamefont{Sheng}},
  \bibinfo{author}{\bibfnamefont{C.-K.} \bibnamefont{Tian}},
  \bibinfo{author}{\bibfnamefont{Y.-Y.} \bibnamefont{Wang}},
  \bibinfo{author}{\bibfnamefont{T.-L.} \bibnamefont{Xia}},
  \bibinfo{author}{\bibfnamefont{L.}~\bibnamefont{Wang}},
  \bibinfo{author}{\bibfnamefont{F.}~\bibnamefont{Ye}},
  \bibinfo{author}{\bibfnamefont{W.}~\bibnamefont{Tian}},
  \bibinfo{author}{\bibfnamefont{J.-C.} \bibnamefont{Wang}},
  \bibinfo{author}{\bibfnamefont{J.-J.} \bibnamefont{Liu}},
  \bibnamefont{et~al.}, \bibinfo{journal}{Euro. Phys. Lett.}
  \textbf{\bibinfo{volume}{122}}, \bibinfo{pages}{67006}
  (\bibinfo{year}{2018}),
  \urlprefix\url{https://dx.doi.org/10.1209/0295-5075/122/67006}.

\bibitem[{\citenamefont{Sefat et~al.}(2019)\citenamefont{Sefat, Nguyen, Parker,
  Fu, Zou, Li, Cao, Sanjeewa, Li, and Gai}}]{12}
\bibinfo{author}{\bibfnamefont{A.~S.} \bibnamefont{Sefat}},
  \bibinfo{author}{\bibfnamefont{G.~D.} \bibnamefont{Nguyen}},
  \bibinfo{author}{\bibfnamefont{D.~S.} \bibnamefont{Parker}},
  \bibinfo{author}{\bibfnamefont{M.~M.} \bibnamefont{Fu}},
  \bibinfo{author}{\bibfnamefont{Q.}~\bibnamefont{Zou}},
  \bibinfo{author}{\bibfnamefont{A.-P.} \bibnamefont{Li}},
  \bibinfo{author}{\bibfnamefont{H.~B.} \bibnamefont{Cao}},
  \bibinfo{author}{\bibfnamefont{L.~D.} \bibnamefont{Sanjeewa}},
  \bibinfo{author}{\bibfnamefont{L.}~\bibnamefont{Li}}, \bibnamefont{and}
  \bibinfo{author}{\bibfnamefont{Z.}~\bibnamefont{Gai}},
  \bibinfo{journal}{Phys. Rev. B} \textbf{\bibinfo{volume}{100}},
  \bibinfo{pages}{104525} (\bibinfo{year}{2019}),
  \urlprefix\url{https://link.aps.org/doi/10.1103/PhysRevB.100.104525}.

\bibitem[{\citenamefont{Cheng et~al.}(2016)\citenamefont{Cheng, Zhang, Bao,
  Schneidewind, Link, Gr¨¹nwald, Georgii, Hao, and Liu}}]{13}
\bibinfo{author}{\bibfnamefont{P.}~\bibnamefont{Cheng}},
  \bibinfo{author}{\bibfnamefont{H.}~\bibnamefont{Zhang}},
  \bibinfo{author}{\bibfnamefont{W.}~\bibnamefont{Bao}},
  \bibinfo{author}{\bibfnamefont{A.}~\bibnamefont{Schneidewind}},
  \bibinfo{author}{\bibfnamefont{P.}~\bibnamefont{Link}},
  \bibinfo{author}{\bibfnamefont{A.}~\bibnamefont{Gr¨¹nwald}},
  \bibinfo{author}{\bibfnamefont{R.}~\bibnamefont{Georgii}},
  \bibinfo{author}{\bibfnamefont{L.~J.} \bibnamefont{Hao}}, \bibnamefont{and}
  \bibinfo{author}{\bibfnamefont{Y.~T.} \bibnamefont{Liu}},
  \bibinfo{journal}{Nucl. Instrum. Methods A} \textbf{\bibinfo{volume}{821}},
  \bibinfo{pages}{17} (\bibinfo{year}{2016}),
  \urlprefix\url{https://dx.doi.org/10.1016/j.nima.2016.03.045}.

\bibitem[{\citenamefont{Xing et~al.}(2016)\citenamefont{Xing, Shi, Richard,
  Wang, Liu, Lv, Ma, Fu, Kong, Miao et~al.}}]{14}
\bibinfo{author}{\bibfnamefont{L.~Y.} \bibnamefont{Xing}},
  \bibinfo{author}{\bibfnamefont{X.}~\bibnamefont{Shi}},
  \bibinfo{author}{\bibfnamefont{P.}~\bibnamefont{Richard}},
  \bibinfo{author}{\bibfnamefont{X.~C.} \bibnamefont{Wang}},
  \bibinfo{author}{\bibfnamefont{Q.~Q.} \bibnamefont{Liu}},
  \bibinfo{author}{\bibfnamefont{B.~Q.} \bibnamefont{Lv}},
  \bibinfo{author}{\bibfnamefont{J.-Z.} \bibnamefont{Ma}},
  \bibinfo{author}{\bibfnamefont{B.~B.} \bibnamefont{Fu}},
  \bibinfo{author}{\bibfnamefont{L.-Y.} \bibnamefont{Kong}},
  \bibinfo{author}{\bibfnamefont{H.}~\bibnamefont{Miao}}, \bibnamefont{et~al.},
  \bibinfo{journal}{Phys. Rev. B} \textbf{\bibinfo{volume}{94}},
  \bibinfo{pages}{094524} (\bibinfo{year}{2016}),
  \urlprefix\url{https://link.aps.org/doi/10.1103/PhysRevB.94.094524}.

\bibitem[{\citenamefont{Zhang et~al.}(2014)\citenamefont{Zhang, Gong, Lu, Li,
  Dai, and Luo}}]{15}
\bibinfo{author}{\bibfnamefont{R.}~\bibnamefont{Zhang}},
  \bibinfo{author}{\bibfnamefont{D.}~\bibnamefont{Gong}},
  \bibinfo{author}{\bibfnamefont{X.}~\bibnamefont{Lu}},
  \bibinfo{author}{\bibfnamefont{S.}~\bibnamefont{Li}},
  \bibinfo{author}{\bibfnamefont{P.}~\bibnamefont{Dai}}, \bibnamefont{and}
  \bibinfo{author}{\bibfnamefont{H.}~\bibnamefont{Luo}},
  \bibinfo{journal}{Supercond. Sci. Technol.} \textbf{\bibinfo{volume}{27}},
  \bibinfo{pages}{115003} (\bibinfo{year}{2014}),
  \urlprefix\url{https://doi.org/10.1088%2F0953-2048%2F27%2F11%2F115003}.

\bibitem[{\citenamefont{Cheng et~al.}(2010)\citenamefont{Cheng, Shen, Hu, and
  Wen}}]{16}
\bibinfo{author}{\bibfnamefont{P.}~\bibnamefont{Cheng}},
  \bibinfo{author}{\bibfnamefont{B.}~\bibnamefont{Shen}},
  \bibinfo{author}{\bibfnamefont{J.}~\bibnamefont{Hu}}, \bibnamefont{and}
  \bibinfo{author}{\bibfnamefont{H.-H.} \bibnamefont{Wen}},
  \bibinfo{journal}{Phys. Rev. B} \textbf{\bibinfo{volume}{81}},
  \bibinfo{pages}{174529} (\bibinfo{year}{2010}),
  \urlprefix\url{https://link.aps.org/doi/10.1103/PhysRevB.81.174529}.

\bibitem[{\citenamefont{Ni et~al.}(2008)\citenamefont{Ni, Tillman, Yan,
  Kracher, Hannahs, Bud'ko, and Canfield}}]{17}
\bibinfo{author}{\bibfnamefont{N.}~\bibnamefont{Ni}},
  \bibinfo{author}{\bibfnamefont{M.~E.} \bibnamefont{Tillman}},
  \bibinfo{author}{\bibfnamefont{J.-Q.} \bibnamefont{Yan}},
  \bibinfo{author}{\bibfnamefont{A.}~\bibnamefont{Kracher}},
  \bibinfo{author}{\bibfnamefont{S.~T.} \bibnamefont{Hannahs}},
  \bibinfo{author}{\bibfnamefont{S.~L.} \bibnamefont{Bud'ko}},
  \bibnamefont{and} \bibinfo{author}{\bibfnamefont{P.~C.}
  \bibnamefont{Canfield}}, \bibinfo{journal}{Phys. Rev. B}
  \textbf{\bibinfo{volume}{78}}, \bibinfo{pages}{214515}
  (\bibinfo{year}{2008}),
  \urlprefix\url{https://link.aps.org/doi/10.1103/PhysRevB.78.214515}.

\bibitem[{\citenamefont{Gong et~al.}(2018)\citenamefont{Gong, Xie, Zhang, Birk,
  Niedermayer, Han, Lapidus, Dai, Li, and Luo}}]{18}
\bibinfo{author}{\bibfnamefont{D.}~\bibnamefont{Gong}},
  \bibinfo{author}{\bibfnamefont{T.}~\bibnamefont{Xie}},
  \bibinfo{author}{\bibfnamefont{R.}~\bibnamefont{Zhang}},
  \bibinfo{author}{\bibfnamefont{J.}~\bibnamefont{Birk}},
  \bibinfo{author}{\bibfnamefont{C.}~\bibnamefont{Niedermayer}},
  \bibinfo{author}{\bibfnamefont{F.}~\bibnamefont{Han}},
  \bibinfo{author}{\bibfnamefont{S.~H.} \bibnamefont{Lapidus}},
  \bibinfo{author}{\bibfnamefont{P.}~\bibnamefont{Dai}},
  \bibinfo{author}{\bibfnamefont{S.}~\bibnamefont{Li}}, \bibnamefont{and}
  \bibinfo{author}{\bibfnamefont{H.}~\bibnamefont{Luo}},
  \bibinfo{journal}{Phys. Rev. B} \textbf{\bibinfo{volume}{98}},
  \bibinfo{pages}{014512} (\bibinfo{year}{2018}),
  \urlprefix\url{https://link.aps.org/doi/10.1103/PhysRevB.98.014512}.

\bibitem[{\citenamefont{Zinth~V and D}(2011)}]{19}
\bibinfo{author}{\bibfnamefont{K.~H.} \bibnamefont{Zinth~V},
  \bibfnamefont{Dellmann~T}} \bibnamefont{and}
  \bibinfo{author}{\bibfnamefont{J.}~\bibnamefont{D}}, \bibinfo{journal}{Angew.
  Chem. Int. Ed.} \textbf{\bibinfo{volume}{50}}, \bibinfo{pages}{7919}
  (\bibinfo{year}{2011}),
  \urlprefix\url{https://dx.doi.org/10.1002/anie.201102866}.

\bibitem[{\citenamefont{Griffiths}(1969)}]{20}
\bibinfo{author}{\bibfnamefont{R.~B.} \bibnamefont{Griffiths}},
  \bibinfo{journal}{Phys. Rev. Lett.} \textbf{\bibinfo{volume}{23}},
  \bibinfo{pages}{17} (\bibinfo{year}{1969}),
  \urlprefix\url{https://link.aps.org/doi/10.1103/PhysRevLett.23.17}.

\bibitem[{\citenamefont{Inosov et~al.}(2013)\citenamefont{Inosov, Friemel,
  Park, Walters, Texier, Laplace, Bobroff, Hinkov, Sun, Liu et~al.}}]{21}
\bibinfo{author}{\bibfnamefont{D.~S.} \bibnamefont{Inosov}},
  \bibinfo{author}{\bibfnamefont{G.}~\bibnamefont{Friemel}},
  \bibinfo{author}{\bibfnamefont{J.~T.} \bibnamefont{Park}},
  \bibinfo{author}{\bibfnamefont{A.~C.} \bibnamefont{Walters}},
  \bibinfo{author}{\bibfnamefont{Y.}~\bibnamefont{Texier}},
  \bibinfo{author}{\bibfnamefont{Y.}~\bibnamefont{Laplace}},
  \bibinfo{author}{\bibfnamefont{J.}~\bibnamefont{Bobroff}},
  \bibinfo{author}{\bibfnamefont{V.}~\bibnamefont{Hinkov}},
  \bibinfo{author}{\bibfnamefont{D.~L.} \bibnamefont{Sun}},
  \bibinfo{author}{\bibfnamefont{Y.}~\bibnamefont{Liu}}, \bibnamefont{et~al.},
  \bibinfo{journal}{Phys. Rev. B} \textbf{\bibinfo{volume}{87}},
  \bibinfo{pages}{224425} (\bibinfo{year}{2013}),
  \urlprefix\url{https://link.aps.org/doi/10.1103/PhysRevB.87.224425}.

\bibitem[{\citenamefont{Gastiasoro and Andersen}(2014)}]{22}
\bibinfo{author}{\bibfnamefont{M.~N.} \bibnamefont{Gastiasoro}}
  \bibnamefont{and} \bibinfo{author}{\bibfnamefont{B.~M.}
  \bibnamefont{Andersen}}, \bibinfo{journal}{Phys. Rev. Lett.}
  \textbf{\bibinfo{volume}{113}}, \bibinfo{pages}{067002}
  (\bibinfo{year}{2014}),
  \urlprefix\url{https://link.aps.org/doi/10.1103/PhysRevLett.113.067002}.

\bibitem[{\citenamefont{Bao et~al.}(2009)\citenamefont{Bao, Qiu, Huang, Green,
  Zajdel, Fitzsimmons, Zhernenkov, Chang, Fang, Qian et~al.}}]{23}
\bibinfo{author}{\bibfnamefont{W.}~\bibnamefont{Bao}},
  \bibinfo{author}{\bibfnamefont{Y.}~\bibnamefont{Qiu}},
  \bibinfo{author}{\bibfnamefont{Q.}~\bibnamefont{Huang}},
  \bibinfo{author}{\bibfnamefont{M.~A.} \bibnamefont{Green}},
  \bibinfo{author}{\bibfnamefont{P.}~\bibnamefont{Zajdel}},
  \bibinfo{author}{\bibfnamefont{M.~R.} \bibnamefont{Fitzsimmons}},
  \bibinfo{author}{\bibfnamefont{M.}~\bibnamefont{Zhernenkov}},
  \bibinfo{author}{\bibfnamefont{S.}~\bibnamefont{Chang}},
  \bibinfo{author}{\bibfnamefont{M.}~\bibnamefont{Fang}},
  \bibinfo{author}{\bibfnamefont{B.}~\bibnamefont{Qian}}, \bibnamefont{et~al.},
  \bibinfo{journal}{Phys. Rev. Lett.} \textbf{\bibinfo{volume}{102}},
  \bibinfo{pages}{247001} (\bibinfo{year}{2009}),
  \urlprefix\url{https://link.aps.org/doi/10.1103/PhysRevLett.102.247001}.

\bibitem[{\citenamefont{Liu et~al.}(2010)\citenamefont{Liu, Hu, Qian, Fobes,
  Mao, Bao, Reehuis, Kimber, Prokes, Matas et~al.}}]{24}
\bibinfo{author}{\bibfnamefont{T.~J.} \bibnamefont{Liu}},
  \bibinfo{author}{\bibfnamefont{J.}~\bibnamefont{Hu}},
  \bibinfo{author}{\bibfnamefont{B.}~\bibnamefont{Qian}},
  \bibinfo{author}{\bibfnamefont{D.}~\bibnamefont{Fobes}},
  \bibinfo{author}{\bibfnamefont{Z.~Q.} \bibnamefont{Mao}},
  \bibinfo{author}{\bibfnamefont{W.}~\bibnamefont{Bao}},
  \bibinfo{author}{\bibfnamefont{M.}~\bibnamefont{Reehuis}},
  \bibinfo{author}{\bibfnamefont{S.~A.~J.} \bibnamefont{Kimber}},
  \bibinfo{author}{\bibfnamefont{K.}~\bibnamefont{Prokes}},
  \bibinfo{author}{\bibfnamefont{S.}~\bibnamefont{Matas}},
  \bibnamefont{et~al.}, \bibinfo{journal}{Nature Materials}
  \textbf{\bibinfo{volume}{9}}, \bibinfo{pages}{718} (\bibinfo{year}{2010}).

\bibitem[{\citenamefont{Thampy et~al.}(2012)\citenamefont{Thampy, Kang,
  Rodriguez-Rivera, Bao, Savici, Hu, Liu, Qian, Fobes, Mao et~al.}}]{25}
\bibinfo{author}{\bibfnamefont{V.}~\bibnamefont{Thampy}},
  \bibinfo{author}{\bibfnamefont{J.}~\bibnamefont{Kang}},
  \bibinfo{author}{\bibfnamefont{J.~A.} \bibnamefont{Rodriguez-Rivera}},
  \bibinfo{author}{\bibfnamefont{W.}~\bibnamefont{Bao}},
  \bibinfo{author}{\bibfnamefont{A.~T.} \bibnamefont{Savici}},
  \bibinfo{author}{\bibfnamefont{J.}~\bibnamefont{Hu}},
  \bibinfo{author}{\bibfnamefont{T.~J.} \bibnamefont{Liu}},
  \bibinfo{author}{\bibfnamefont{B.}~\bibnamefont{Qian}},
  \bibinfo{author}{\bibfnamefont{D.}~\bibnamefont{Fobes}},
  \bibinfo{author}{\bibfnamefont{Z.~Q.} \bibnamefont{Mao}},
  \bibnamefont{et~al.}, \bibinfo{journal}{Phys. Rev. Lett.}
  \textbf{\bibinfo{volume}{108}}, \bibinfo{pages}{107002}
  (\bibinfo{year}{2012}),
  \urlprefix\url{https://link.aps.org/doi/10.1103/PhysRevLett.108.107002}.

\end{thebibliography}
\end{document}